\documentclass[prl,floatfix,showpacs,twocolumn,preprintnumbers]{revtex4}
\usepackage{amsmath,amssymb}
\usepackage{graphicx,color}

\begin{document}
\title{Critical properties  of a continuous family  of XY noncollinear
magnets}

\author{A. Peles$^{\dagger}$,  B.W. Southern$^{\ddagger}$,  B.
Delamotte$^{\star}$, D.  Mouhanna$^{\star}$ and  M.
Tissier$^{\diamond}$}
\address{$^{\dagger}$Department of Physics and
     Astronomy, University of Manitoba,\\ Winnipeg Manitoba, Canada R3T
     2N2 and \\  School of Physics, Georgia Institute of Technology, Atlanta, Georgia 30332-0430\\
$^{\ddagger}$Department of Physics and
     Astronomy, University of Manitoba,\\ Winnipeg Manitoba, Canada R3T
     2N2\\ $^{\star}$Laboratoire de Physique Th\'eorique et Hautes Energies,
      Universit\'es Paris VI-Pierre et Marie Curie - Paris
     VII-Denis Diderot\\ 2 Place Jussieu, 75251 Paris Cedex 05 France.\\
$^{\diamond}$Laboratoire de Physique Th\'eorique des Liquides\\
      Universit\'e Paris VI-Pierre et Marie Curie\\ 4 Place
     Jussieu 75252 Paris Cedex 05 France.}

\date{\today}

\begin{abstract}
     Monte Carlo methods are used to study a family of three
     dimensional XY frustrated models interpolating continuously between
     the stacked triangular antiferromagnets and a variant of this model
     for which a local rigidity constraint is imposed.  Our study leads us
     to conclude that generically weak first order behavior occurs in
this family of
     models in agreement with a recent nonperturbative renormalization group
     description of
     frustrated magnets.
\end{abstract}


\maketitle

In spite of intensive study during the last 25  years the
critical behavior of XY or Heisenberg frustrated magnets with a
noncollinear ground state is still a strongly debated topic
(see Refs. \cite{delamotte03,kawamura98} and references therein).
This is, for instance, the case of the celebrated XY Stacked
Triangular Antiferromagnet (STA) that we consider here, whose
Hamiltonian is:
\begin{equation}
H=-\sum_{\langle ij\rangle}J_{ij}\, {\bf S}_i. {\bf S}_j
\label{hmicro}
\end{equation}
where the ${\bf S}_i$ are two-component vectors and
the sum runs over all pairs of nearest neighbor spins on a stacked
triangular lattice.
   In Eq.~(\ref{hmicro}), the in-plane interactions are antiferromagnetic
($J<0$) and the inter-plane interactions are taken to be
ferromagnetic ($J>0$) with
$|J|=1$. The competition between the
in-plane antiferromagnetic interactions produces a ground state where
the three spins ${\bf S}_1, {\bf S}_2$ and ${\bf S}_3$ on each
elementary  triangular  plaquette are oriented at 120$^{\circ}$ one
to another,  that is:
\begin{equation}
{\bf S}_1+{\bf S}_2+{\bf S}_3={\bf 0}.
\label{rigidity}
\end{equation}
For such a system, the order parameter is  given by two orthogonal
vectors of the same norm, a fact that  has led  to the hypothesis that
noncollinear magnets could undergo a second order phase transition
associated with a new ---  chiral --- universality
class~\cite{kawamura85}.
 
However, it is now becoming more and more widely accepted that
    the physics of frustrated magnets is not  given  by such a
simple picture~\cite{tissier01,delamotte03,calabrese02,calabrese03c}.
The complexity of the critical behavior of STA systems is
exemplified in the experimental results (for a review
see Ref. \cite{delamotte03} and references therein). Indeed, scaling
behavior is generally found while universality is violated since
materials which belong, {\it a priori}, to the same universality class
display different critical exponents. It is thus difficult to
interpret the experimental data within the usual picture of critical
phenomena.
 
On the theoretical side, the situation is
controversial~\cite{delamotte03,calabrese02}.
A nonperturbative renormalization group (NPRG)
method~\cite{tissier01,delamotte03} finds that the phase
transitions in these systems are {\it generically}
weakly  first order with the possibility of standard strongly first order
transitions.  More precisely while there is no RG  fixed point  there
exists a  whole region  where the RG flow is very slow,
corresponding  to large --- but finite --- correlation lengths  at
the transition,
which allows one to compute pseudo-critical exponents. Hence the violation of
universality observed experimentally has a natural explanation since,
in the absence of a fixed point, the pseudo-critical exponents
found  depend on the microscopic Hamiltonians.
On the other hand, Pelissetto {\it et al.} have derived and resummed
the six-loop  $\beta$ functions  in three dimensions~\cite{pelissetto01a}.
They find a fixed point and therefore predict a second order phase transition.
Calabrese {\it et al.}
have claimed that the lack of universality found can  be accounted for by the
spiral  nature of the
approach to the {\it focus} fixed point~\cite{calabrese02}.

Our aim is to discriminate  by means of numerical simulations between
these two scenarios that
both predict scaling behavior with varying critical exponents  but
differ in the predicted asymptotic behavior for temperatures very
close to the transition temperature.
Different systems having the same symmetry breaking scheme
as XY-STA but differing by microscopic details have been studied
numerically~\cite{diep89,kunz93,loison98,
itakura03,kawamura86,kawamura92,plumer94,boubcheur96}. Apart from STA,
they have {\it all}  been found to undergo first order
transitions~\cite{diep89,kunz93,loison98,itakura03}. As for STA itself
early numerical simulations favor a continuous
transition~\cite{kawamura86,kawamura92,plumer94,boubcheur96}. However
the most recent simulation performed with lattice sizes much larger
than those considered previously leads to  first order
behavior~\cite{itakura03}. Thus, although there are now strong
indications in favor of the nonperturbative scenario, the numerical
evidence is still too weak to lead to a definite conclusion. In order to shed
some light on this problem we consider a family of
models, built up as generalizations of STA. We find  that they {\it
all} exhibit either
strong first order transitions or a scaling behavior with varying
exponents associated with a weak first order phase transition.

To perform our study we partition the original triangular lattice ---
with lattice spacing $a$ --- into elementary plaquettes labelled by an
index $I$. They are composed of three spins ${\bf S}_1^I,{\bf S} _2^I$
and ${\bf S}_3^I$. The super lattice made of these plaquettes is,
again, a triangular lattice, with lattice spacing $\sqrt 3 a$. The
family of models we consider is defined by:
\begin{equation}
H(r)=-\sum_{\langle ij\rangle}J_{ij}\, {\bf S}_i. {\bf
    S}_j +\, r \hspace{0.1cm}\sum_{\hbox{\scriptsize I}}\hspace{0.2cm}
({\bf S}_1^I+{\bf S}_2^I +{\bf S}_3^I)^2\
\label{hstastar}
\end{equation}
where the first term is the same as in Eq.~(\ref{hmicro}) and where
the sum over $I$ in the second $r$-term runs over the indices of the
super lattice.  For $r < -0.5 |J|$, the ground state has the spins
aligned ferromagnetically within a
    plaquette but  are oriented at $120^\circ$ from plaquette to
    plaquette; this is not the ground state expected for real materials.
    For $r=-0.5 |J|$ the interactions inside the plaquettes vanish and
    the lattice becomes effectively a stacked Kagome lattice.  For  $r > -0.5
    |J|$ the ground state and thus the symmetry breaking scheme and
order parameter are identical to those of the STA model,
Eq.(\ref{hmicro}); we thus focus on this range of values in the
    following.  For any value of $r>0$ the $r$-term favors locally,
that is {\it inside} the plaquettes, the $120^\circ$ ground state
configuration. It thus penalizes the relative fluctuations of the
spins ${\bf S}_1^I, {\bf S}_2^I$ and ${\bf S}_3^I$ inside a plaquette
but still lets the orientation of the three-spin structure freely
fluctuate from plaquette to plaquette.  Note that, since the
fluctuations of the spins inside a plaquette are noncritical, the
$r$-term should not alter the critical physics of the STA {\it if} it
is universal --- see below. The $r\to\infty$ limit consists of a STA
for which the ground state 120$^{\circ}$ structure is locally imposed
at {\it all} temperatures. This model is called the STAR (R for
rigid). Continuous changes in the `rigidity parameter' $r$ from zero
to infinity thus correspond to a continuous change from the STA to the
STAR. The STAR has been studied by Loison and Schotte~\cite{loison98}
and has been found to undergo a strong first order transition. This
already shows that, in fact, the critical behaviors of STA-like
systems strongly depend on the microscopic details of the model.

We use a classical Monte Carlo Metropolis algorithm~\cite{metropolis}
to study this generalized model as a function of temperature and the
rigidity parameter $r$. A number of different thermodynamic quantities
have been computed with $ 0 \leq r \leq 8 $
for lattices of linear
sizes $L=18$ to $L=138$.  In order to reduce the critical
slowing down at the transition, each simulation run starts from a
disordered state at a temperature well above the transition temperature
and is slowly cooled to low temperatures.  The system is allowed to
equilibrate for $3 \times 10^{5}$ Monte Carlo sweeps (MCS) and the averages
were performed over $5.5 \times 10^{5}  $ steps. In order to avoid
correlations between subsequent measurements, $20 -30$ Monte Carlo
sweeps were performed between measurements.

\begin{figure}[t]
\centering
\includegraphics[width=3.5in,angle=0]{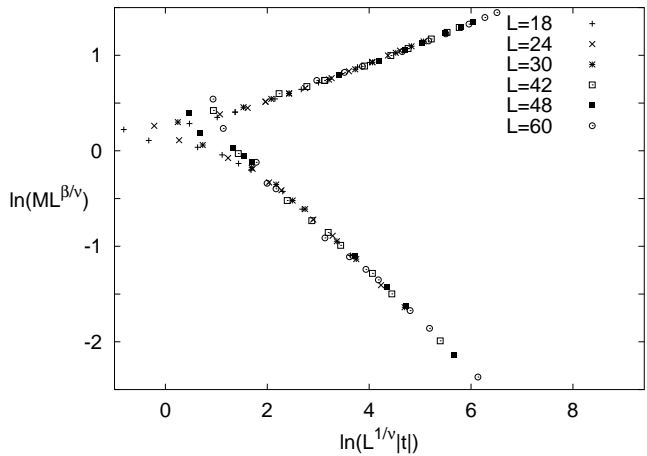}
\caption{Log-log finite size scaling plot of order parameter $M$ for
the $r=1.0$ system.
The transition temperature is $T^* = 1.7369$ and the exponent  $\beta
=0.2134$. The
    best collapse of data
was obtained for $\nu=0.4817$ .}
\label{ch6_fig2}
\end{figure}

{\it $0 \leq r \leq 1.0$: the weak first order region.}  We begin our
discussion with  the small $r$ region
   which is the most interesting
one since real compounds are supposed to be close to STA. Moreover,
for some of these compounds as well as for STA with $r=0$ scaling behavior is
found \cite{delamotte03}. Since scaling laws are found to hold in
this range of $r$, we
   analyze them
by means of the standard methodology used for second order phase
transitions. The fourth order cumulant $U_M$ of the
order parameter displays behavior characteristic of a second order
transition and can be used to estimate the transition temperature $T^*$.
We have determined  $T^*$, the order
parameter critical exponent $\beta$, the correlation length exponent
$\nu$, the susceptibility exponent $\gamma$ and the anomalous
dimension exponent $\eta$ using  finite size scaling  (FSS)  methods 
\cite{kawamura92}.
For appropriate values of $T^*$ and of  $\beta$ and $\nu$, the data should
collapse onto a
single curve. A sample log-log  finite size scaling plot of the order
parameter $M$ for $r=1.0$ is shown in figure
\ref{ch6_fig2}.   A similar FSS analysis of the susceptibility also
satisfies scaling and can be used to determine  $\gamma$.
For sizes $L=48, 60$  there are some deviations from scaling at small
values of the
reduced temperature $t$.

\begin{table}[!tbp]
\begin{center}
\begin{tabular}{|c|c|c|c|c|c|}
\hline
$r$&$T^*$  & $\beta$ & $\gamma$ & $\nu$ & $\eta$\\
\hline
0.0&1.458(2)~ &~~~0.253(10)&~1.13(5)~&~~~0.54(2)~~~&-\\
\hline
0.2&1.5494(4) &~~0.249(6)~&~1.1(1)~~&~~~0.531(20)&~~~-0.06(2) \\
\hline
0.5&1.6420(3)&~~0.243(9)~&~~1.04(10)&~~~0.518(20) &~~~-0.08(4) \\
\hline
0.8&1.7039(7) &~~~0.218(10)~&~1.0(1)~~&~~~0.490(20)&~~~-0.04(3)\\
\hline
1.0&1.737(1)~ &~~0.213(9)~&~1.0(2)~~&~~~0.482(25)&~~~-0.07(3)\\
\hline
\end{tabular}
\caption{Critical exponents for various $r$.
The results for $r=0$ were obtained in \cite{kawamura92}.}
\label{tabb}
\end{center}
\end{table}
At $T^*$, the structure factor depends on system size as
$S(\vec{Q},L) \sim L ^{2-\eta}$ where the exponent $\eta$ is the
anomalous
dimension and $\vec{Q}$ is the ordering wave vector for the three
sublattice structure.
Throughout the entire range $0\le r <1.0$,  we again find that
scaling is satisfied  for
system sizes $L \le 42$. Table~\ref{tabb} summarizes our results for
the critical exponents
in the range $0\le r\le1.0$.   The critical exponents for which
finite size scaling holds change with  $r$ showing clearly that
universality is violated. Moreover $\eta$ is always found to be
negative --- sometimes significantly, e.g. for $r=0.2$ --- which is
impossible if the transition is of second order \cite{delamotte03}.
These results strongly suggest that scaling aborts ultimately ---
that is, very close to $T^*$ ---  for all $r\in [0,1]$ and that,
therefore, the transitions are {\it all} of first order.

{\it $r\ge$ 1.0:  the strong first order region.} We now consider
the  region  $r\ge1.0$. As expected, the first order  behavior found
for the STAR, that is for  $r\to\infty$, persists for finite $r$. We
have been able to follow it down  to $r\simeq 1.0$
   by studying  the probability distribution for the energy $P(E)$ .
This probability distribution
is a useful quantity  to locate first order
transitions. Away from the  transition point,
one expects a Gaussian distribution in energy while at the
     transition point a double peak in $P(E)$ occurs if it is first order.
    This point can be located by finding the
temperature where the weights under the peaks in $P(E)$ become equal.
However distinguishing between a weak first order and a
second order transition is rather delicate especially in case of a
possible tricritical point.  In order to make sure what type of phase
transition the double peak structure represents, it is necessary to
study the size dependence of the probability distribution $P(E)$.  In the
case of a first order transition, the separation of the peak energies 
$E_1$ and $E_2$  equals the latent heat up to $1/L$ corrections 
\cite{lee}. Thus two peaks  remain separated with increased  size $L$ 
and, as shown in \cite{binder2}, become sharper.

We have found a double peaked structure in $P(E)$  for rigidity
parameters $r=1.0,1.5,2.0,4.0,6.0$ and $8.0$ for linear sizes $L \le 60$.
As the value of $r$
   decreases, the size of the system where the double peak
structure starts to appear increases.  However, the energy autocorrelation time
   is less than $2000$ MCS in this range.
The double peaked structure in $P(E)$, and thus the first
order behavior, is apparent but only for lattice sizes $L > 48$ for 
$r=1.0$.  If we extrapolate to the
thermodynamic limit, we can estimate the latent heat $\Delta E = |E_1
- E_2|$ at the transition.  Using the values of $\Delta E$ for $r \ge 1.0$
and extrapolating to the more interesting small $r$ region predicts
that $\Delta E$ remains nonzero at $r=0$ with an estimated value of
$0.010 \pm 0.002$. This
behavior suggests that the phase transition at $r=0$ might be weakly
first order.

\begin{figure}[t]
\centering
\includegraphics[width=3.5in,angle=0]{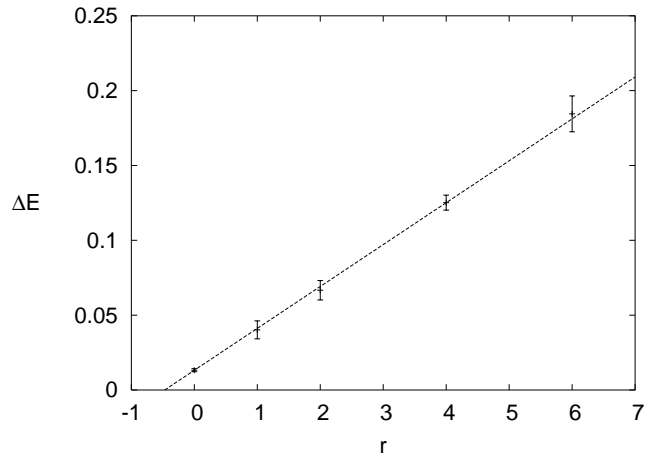}
\caption{The dashed line corresponds to a linear fit of the energy
difference of peak maxima
for $r=1, 2, 4$ and $6$.  The point at $r=0$ was estimated from the 
histograms in figure 3.}
\label{deltaer}
\end{figure}

  In figure \ref{deltaer} we plot $\Delta E $
vs $r$ where it can be seen that it extrapolates linearly to the physically
interesting limit of small $r$. The maximum size used for $r=6$ was
$L=36$ and hence the
errorbars are larger than those at smaller $r$. The value at $r=0$ is
finite indicating a weak first order
transition.  The extrapolation also
predicts that $\Delta E$ vanishes at $r_0=-0.47 \pm 0.02$.  Our error
bars for $r_0$ are compatible with $r=-0.5$,
    corresponding to the stacked Kagome lattice, for which  one 
expects a completely different physics.
We conclude that STA undergoes a very weak first order phase
transition and that there is no tricritical point for all systems
described by $H(r)$   with $r > -0.5$.

\begin{figure}[b]
\centering
\includegraphics[width=3.5in,angle=0]{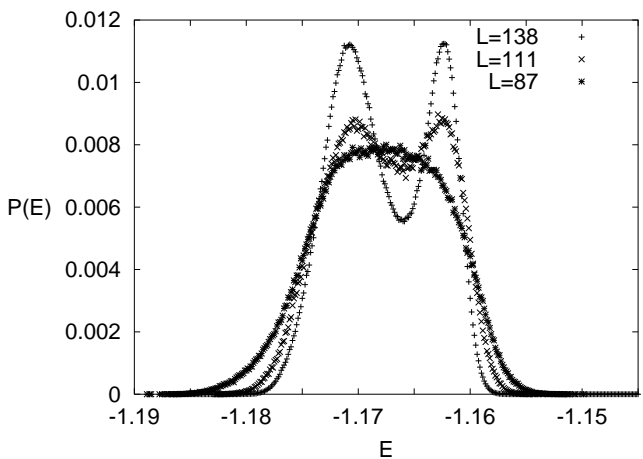}
\caption{P(E) for STA $XY$ model ($r=0$) for $L=87, 111$ and $L=138$ at $T^*$.}
\label{R0.0peaks}
\end{figure}

{\it STA $r=0$}.
In order to confirm our result of a non-zero latent heat at $r=0$,
we have studied  the energy probability distribution  for  the $r=0$
case using much larger system  sizes. Figure \ref{R0.0peaks} shows
$P(E)$ at $T^*$
for  sizes $L=87, 111$ and $138$. A double peak structure characteristic of a
first order phase transition is now evident at these sizes.
As the size of the system is increased, the peaks sharpen and  the
depth of the minimum between the peaks also increases.  For these larger
sizes the energy autocorrelation time increases significantly to
values greater than $12000$ MCS.
This critical slowing down could perhaps be improved
using overrelaxation or multicanonical histogram methods.\cite{binder2}
However, we estimate a value  $\Delta E \sim 0.013 \pm .001$   from figure
\ref{R0.0peaks} which is in excellent agreement with the previous
estimate $0.010 \pm 0.002$
obtained from the extrapolation of $\Delta E (r)$ from larger $r \ge
1.0$ to $r=0$.

In conclusion, we have studied a system interpolating continuously
between the $XY$ STA and STAR models and find unambiguous evidence
that the phase transitions in these systems is either of first order
or weakly first order accompanied by scaling with exponents varying
with $r$.  These results provide strong arguments in favor of the
nonperturbative approach to the physics of frustrated
magnets~\cite{delamotte03} and  suggest  that, even if a fixed
point exists, its relevance for STA-like compounds whose microscopic
Hamiltonians are supposed to be well described by Eq.~(\ref{hmicro}),
is doubtful. Indeed, while the two --- perturbative and
nonperturbative --- approaches both predict scaling in a transient
regime near $T^*$, they differ in several ways. First, asymptotically,
{\it i.e.}  at $T^*$ --- the scaling behavior aborts in the
nonperturbative scenario while it persists in the perturbative one
since it is associated with a true continuous phase transition.
Second, the weak first order transitions are generic in the
nonperturbative scenario whereas they should be exceptional in the
perturbative one~\cite{delamotte03}. Indeed, in the presence of a
fixed point, systems undergoing weak first order phase transitions
correspond, in the space of coupling constants, to initial conditions
of the RG flow that are outside the basin of attraction of the fixed
point but that lie in the vicinity of the separatrix between the first
and second order regions. This behavior should not be generic since it
implies a fine-tuning of the parameters of the microscopic
Hamiltonian.  Note however that one cannot completely exclude
    that, by varying another microscopic parameter like the ratio
    between the interlayer and intralayer coupling constants, the system
    be driven into the basin of attraction of a fixed point.

The present work as well as the one performed by
Itakura~\cite{itakura03} predict a first order phase transition for
STA. Previous numerical simulations failed to reach the asymptotic
regime where the correlation length saturates, that is, where the
scaling aborts. This suggests that the same phenomenon could occur in
experiments, this time due to the difficulty to probe the close
vicinity of $T^*$, and that all real materials also display weak first
order transitions. This hypothesis is reinforced by the fact that the
exponents found in our study are close to the experimental ones for
STA compounds (see \cite{delamotte03} and references therein). Let us
consider for instance the exponent $\beta$ which is the best measured
one. Here we find a variation of $\beta$ in the range $[0.213,0.253]$
while for XY STA compounds one has, for instance, $\beta\simeq 0.228$
for CsMnBr$_3$, $\beta\simeq 0.243$ for CsNiCl$_3$ and $\beta\simeq
0.23-0.25$ for CsCuCl$_3$ which has been finally found to undergo a
first order transition when the study of the close vicinity of $T^*$
has been refined.

An important open question concerns the Heisenberg case where there
exists the same conflict between the perturbative and nonperturbative
scenarios.  However, according to the NPRG scenario, the transition
should be even more weakly of first order for Heisenberg than for XY
spins so that the controversy will be more difficult to settle.

This work was supported by the Natural Sciences and Research Council
of Canada, the HPC facility at the
University of Manitoba and HPCVL at Queen's University. Laboratoire
de Physique Th\'eorique et Hautes Energies is UMR 7589. Laboratoire
de Physique
Th\'eorique des Liquides is UMR 7600.

\end{document}